\documentclass[12pt]{article}
\usepackage{amssymb}
\usepackage{amsfonts}
\usepackage{amsmath}
\usepackage{amsmath,amsthm}
\usepackage{subfigure}
\usepackage{graphicx,epsfig, color }

\oddsidemargin 10mm \numberwithin{equation}{section}
\def\be{\begin{equation}}
\def\ee{\end{equation}}
\usepackage[numbers,sort&compress]{natbib}
\usepackage[top=3.5cm,right=3cm,bottom=3cm,left=2cm]{geometry}
\begin{document}
\begin{center}
{{\bf {Thermodynamic Phase Transition of Generalized Ayon-Beato
Garcia Black Holes}} \vskip 1 cm { Elham Ghasemi \footnote{E-mail
address: e\_ ghasemi@semnan.ac.ir} and Hossein Ghaffarnejad
\footnote{E-mail address: hghafarnejad@semnan.ac.ir}
} }\\
\vskip 0.1 cm
{\textit{Faculty of Physics, Semnan University, P.C. 35131-19111, Semnan, Iran} } \\
\end{center}
\begin{abstract}
In this work we study thermodynamics of generalized Ayon-Beato and
Garcia (ABG) black hole metric which contains three parameters
named as mass $m$, magnetic charge $q$ and dimensionless coupling
constant of nonlinear electrodynamics interacting field $\gamma$.
We showed that central regions of this black hole behaves as
dS(AdS) vacuum space by setting $q<2m (q>2m)$ and in the case
$q=2m$ reaches to a flat Minkowski space. In the large distances
this black hole behaves as a Reissner-Nordstrom BH. However
important role of the charge q is appeared in produce of a formal
variable cosmological parameter which will support pressure
coordinate in the thermodynamic perspective of this black hole in
our setup. We should be point that this formal variable
cosmological parameter is different with cosmological constant
which comes from AdS/CFT correspondence and it is effective at
large distances as AdS space pressure.  In our setup the assumed
pressure is originated from internal material of the black hole
say q and m here. By calculating the Hawking temperature of this
black hole we obtain equation of state. Then we plotted isothermal
P-v curves and heat capacity at constant pressure. They show that
the system participates in the small to large phase transition of
the black hole or the Hawking-Page phase transition which is
similar to the Van der Waals phase transition in the ordinary
thermodynamics systems . In fact in the Hawking-Page phase
transition disequilibrium evaporating generalized ABG black hole
reaches to a vacuum AdS space finally.
\end{abstract}
\section{Introduction}
The Einstein general theory of relativity is the most efficient
theory of the gravity, which its validity has been approved
through its correspondence to the observations and experiments in
many years \cite{1}. But in some cases, this theory is not
practical. For instance the spacetime singularity, which is an
example of the failure of general relativity. The gravitational
causal singularity is the extreme density and as a result, so
intense gravity in a point of spacetime where the spacetime breaks
down. These causal singularities  are appeared in metric solutions
of the Einstein`s gravitational field equations and they can not
omitted by usual coordinate transformations.  Penrose-Hawking
singularity theorems show by holding the Einstein`s metric
equations under some circumstances the existence of space time
singularities is unavoidable \cite{2,3,4}. In order to avoid the
central singularity of black holes, Penrose suggested his cosmic
supervision hypothesis which says: the singularity of black hole
is always hidden behind its event horizon \cite{5}. Nevertheless,
many agrees the singularity is generated by classical gravity
theories, while they are neither physical nor exist in universe
\cite{6}. Sakharov \cite{7} and Gliner \cite{8} firstly showed by
considering the effects of quantum, the spacetime singularity is
avoidable. Bardeen \cite{9} inspired by Sakharov`s idea, he
proposed the first singularity free solution of black holes which
are called as regular black holes now. He suggested a static
spherically symmetric solution without considering a known
physical source. Later, numerous different kinds of regular black
holes were suggested \cite{10,11,12,13,14}. Among them, Ayon-Beato
and Garcia (ABG) \cite{15} considered a nonlinear electromagnetic
field as a physical source to produce regular black holes.  In
this way, they obtained ABG regular black holes by solving the
Einstein`s metric equations which are coupled with  suitable
nonlinear electromagnetic fields.  By following this method, other
authors also confirmed ABG regular black holes \cite{16,17,18}.
Cai and Miao \cite{19} achieved a kind of generalized ABG related
black hole solutions which are dependent on five parameters named
mass, charge and three parameters related to nonlinear
electrodynamic fields. This kind of black hole returns to regular
black hole under special conditions. Also, ABG black hole
\cite{15} and its other generalization \cite{20} is obtained under
some assumptions. In \cite{19} a new family of ABG black holes
have been focused which have three parameters named mass, charge
and dimensionless parameter
\textcolor[rgb]{0.00,0.00,0.00}{$\gamma$.} Cai and Miao \cite{19}
studied quasinormal modes and shadows radius for this new family
of ABG black hole and also analyzed the effects of charge and
$\gamma$ parameter on event horizon radius and Hawking
temperature. Hawking by considering quantum effects, showed black
holes radiate like black bodies with particular temperature
\cite{21}, related to surface gravity of black holes horizon and
Bekenstein attributed entropy to black holes which is related to
area of surface of the black holes horizon as $S=A/4$ \cite{22}.
These two discoveries lead us to investigate the thermodynamic
behavior of black holes. Black hole thermodynamics is the
consequence of relation between general relativity and quantum
field theory which guide us to the unknown quantum gravity. By
considering black holes as thermodynamic systems, Bardeen, Carter
and Hawking \cite{23} rewrote the four laws of thermodynamic for
black holes. In this way, Davies studied the phase transition of
Kerr black hole in \cite{24}. In study thermodynamics of the black
holes in usual way we need pressure thermodynamic coordinate which
is bring from cosmological constant in extended phase space with
negative value. In fact this is originate from CFT/AdS
correspondence where one can investigate to study thermodynamics
of the black holes. In this way the Hawking and Page discovered a
first order phase transition for black holes in Anti-de Sitter
(AdS) spacetime \cite{25}. Other types of phase transitions have
been followed in other works \cite{26,27,28,29,30,31}. Since the
cosmological constant has been suggested as thermodynamic pressure
of AdS space background which effects on evaporating quantum black
hole for settings of its thermodynamic equation of state
\cite{32,33,34,35}, the attentions have been attracted to black
hole thermodynamics in extended phase space
\cite{36,37,38,39,40,41,42}. However we will use other idea to
bring a pressure coordinate in studying the generalized ABG black
hole in this work. In fact central region of the ABG black hole
behaves as dS/AdS space where the black hole magnetic charge plays
an important role to produce a formal cosmological parameter and
so to study  thermodynamics of this kind of black hole we do not
need to add cosmological constant which comes from AdS space. In
other words this kind of black hole is self contained while for
some well known black holes for instance Schwarzschild or
Reissner-Nordstrom which are singular at central regions we must
be add an additional cosmological constant parameter to produce
pressure term. Layout of this work is as follows.
\\
In section 2 we define metric of generalized ABG black holes
briefly. In section 3 we investigate thermodynamic perspective of
the model. In section 4 we study possibility of the black hole
phase transition. Section 5 is dedicated to summary and
conclusion.
\section{ Generalized ABG black hole}
Let us we start with the following nonlinear Einstein Maxwell
action functional \cite{15}
\begin{equation}
\label{action}
S=\int d^4x \sqrt{-g} \Big[\frac{R}{16\pi}-\frac{L(P)}{4\pi} \Big]
\end{equation}
in which, $R=g_{\mu\nu}R^{\mu\nu}$ is Ricci scalar and $g=|det
g_{\mu\nu}|$ is absolute value of determinant of metric tensor
field. Nonlinear electromagnetic field lagrangian density
$L(P)=2PH_P-H(P)$ is coupled as minimally with the gravity  where
$P\equiv\frac{1}{4}P_{\mu\nu}P^{\mu\nu}$ is a gauge invariant
scalar. $P_{\mu\nu}\equiv\frac{F_{\mu\nu}}{H_P}$ is nonlinear
antisymmetric tensor versus the electromagnetic tensor field
$F_{\mu\nu}\equiv
\partial_{\mu} A_{\nu}-\partial_{\nu} A_{\mu}$.Here $A_{\mu}$ is electromagnetic
potential and $H_p=\frac{dH(P)}{dP}$ in which $H(P)$ is a
structure function of nonlinear electrodynamic field given by
\cite{19}
\begin{equation}H(P)=\frac{P[1-(\frac{\beta\gamma}{2}-1)(-2Pq^2)^\frac{\gamma}{4}]}{[1+(-2Pq^2)^\frac{\gamma}{4}]^{1+\frac{\beta}{2}}}
-\frac{\alpha\gamma
m(-2Pq^2)^{\frac{3+\gamma}{4}}}{2q^3[1+(-2Pq^2)^{\frac{\gamma}{4}}]^{1+\frac{\alpha}{2}}}.\end{equation}
 By looking at the ref. \cite{19}, one can infer that the
above model has a spherically symmetric static black hole metric
field as,
\begin{equation}
ds^2=-f(r)dt^2+f(r)^{-1}dr^2+r^2(d\theta^2+\sin\theta^2d\varphi^2)
\end{equation}
in which
\begin{equation}
\label{f}
f(r)=1-\frac{2mr^{\frac{\alpha\gamma}{2}-1}}{(q^\gamma+r^\gamma)^{\alpha/2}}+\frac{q^2
r^{\frac{\beta\gamma}{2}-2}}{(q^\gamma+r^\gamma)^{\beta/2}}.
\end{equation}
This is called generalized ABG black hole metric potential for
which $m$ is ADM mass parameter and $q$ is magnetic charge and
three different dimensionless parameters $\alpha$, $\beta$ and
$\gamma$ are associated to nonlinear electrodynamic fields. By
choosing different values for these parameters one can show that
the metric \eqref{f} may become non-singular or naked singular at
central regions \cite{9}. For particular choice
$(\alpha,\beta,\gamma)=(3,4,2)$ the generalized ABG metric field
\eqref{f} returns to original ABG black hole solution \cite{15}
and it goes to other generalized ABG black hole solutions
\cite{20} by setting $\gamma=2$. For simplicity we set here a
particular choice of the above mentioned parameters as
$\alpha\gamma=6$ and $\beta\gamma=8$ for which the generalized ABG
black hole metric given by ref. \cite{19} reduces to the following
form.
\begin{equation}
\label{f1} f(r)=1-\frac{2mr^2}{\left(r^\gamma+q^\gamma \right)
^{3/\gamma}}+\frac{q^2r^2}{\left(r^\gamma+q^\gamma \right)
^{4/\gamma}}.\end{equation} It is easy to check that the above
metric solution reduces to the well known Reissner-Nordstr\"{o}m
form $f(r)\sim1-\frac{2m}{r}+\frac{q^2}{r^2}$ for limits
$\frac{r}{q}\rightarrow\infty$ and for central regions of the ABG
black hole $\frac{r}{q}\to0$ we will have a flat Minkowski form
for $q=2m$ and dS(AdS) $f(r)\sim1+(q-2m)(r^2/q^3)$ for $q<2m
(q>2m)$ respectively. In other words one can infer that for
particular choice $q=2m$ the ABG black hole reads to a flat
Mikowski form at central regions $r<<|q|$ while it reaches to an
extremal RN form at large distance $r>>|q|.$ However, by assuming
that the cosmological parameter in this model is originated just
from ABG magnetic charge $q$ not an AdS background which surrounds
the ABG black hole, therefore we can choose components of the
metric potential as follows.
\begin{equation}\label{Lan}\Lambda(r)=\frac{-3q^2}{(r^\gamma+q^\gamma)^\frac{4}{\gamma}}.\end{equation}
This is in fact a formal variable cosmological parameter and
\begin{equation}\label{Mr}M(r)=\frac{mr^3}{(r^\gamma+q^\gamma)^{\frac{3}{\gamma}}}\end{equation} as distribution of mass function where the
metric potential (\ref{f1}) appears similar to the Schwarzschild AdS black hole form such that
\begin{equation}f(r)=1-\frac{2M(r)}{r}-\frac{1}{3}\Lambda(r)r^2.\end{equation} However if we want to study thermodynamics of this
generalized ABG black hole just near the central regions it is enough to extract a variable cosmological parameter such as above
form or other alternative which is proposed in the next section from metric potential itself (\ref{f1}) because it behaves as dS/AdS
 form just near the central regions of the nonsingular ABG black hole.
 As we mentioned in the abstract of this paper this approach is different with respect to usual way
where a black hole is surrounded by an AdS vacuum space with
constant pressure related to a negative cosmological constant.
This latter proposal comes from AdS/CFT correspondence and the
used cosmological constant comes from geometrical approach and it
should be describes inflation of the universe in the $\Lambda CDM$
model (see for instance \cite{HB}). In other words this is
effective at large distances while (\ref{Lan}) is vanishing at
large distances and has not any relation to the AdS vacuum space.
In fact our assumed variable  pressure (or formal variable
cosmological parameter) originates from internal materials of the
black hole under consideration namely  magnetic ABG  charge q and
the ADM mass m here. By looking at the above definitions one can
obtain easily  $M(0)=0$ and $M(\infty)=m$ with corresponding
values for the cosmological parameter as
$\Lambda(0)=-\frac{3}{q^2}$  and $\Lambda(\infty)=0$ which by
copping to proposal of the AdS/CFT correspondence we can relate
charge of the ABG nonlinear electromagnetic field $q$  to an
assumed central AdS space radius $\ell_{AdS}$ such that
$|q|\sqrt{3}=\ell_{AdS}$. In fact these boundary values for
$\Lambda$ show that central region of the ABG black hole behaves
as AdS space while at large distances the metric is asymptotically
flat. In other words central region of the ABG black hole can be
assumed that is filled with some dark matter with a repeller force
which prevents the central area of the black hole from collapsing.
In this view one can infer that the above assumed variable
cosmological parameter is in fact different with which one  came
from proposal of the AdS/CFT correspondence in the large scales of
bulk spacetime. Because the ansatz   cosmological constant which
is used in the usual AdS/CFT  correspondence generated from
symmetry group between  gauge theory on the boundary and bulk
gravity theory should be vanishing at large cosmological scales
while in our approach it reaches to some small values at central
regions of  the ABG black hole. From this point of view our idea
differ with usual proposal about the unknown cosmological constant
which is used in the general theory of relativity to describe the
inflation with positive value or to adjust the pressure component
in the study of thermodynamics of black holes with a negative sign
(AdS black holes). In the subsequent section we study
thermodynamics of the above ABG black hole but with the above
mentioned perspective about the negative cosmological parameter.
\section{Thermodynamics of generalized ABG black hole}
By regarding the previous perspective about a variable formal
cosmological parameter to study thermodynamics of the ABG black
hole it is useful to assume central region of the ABG  black hole
metric behaves as alone AdS form as
$f(r)=1+\frac{\Lambda(r)r^2}{3}$  in which  $\Lambda(r)$ is
alternative variable cosmological parameter such that
\begin{equation}\Lambda(r)=-8\pi P(r)=3\bigg[\frac{q^2}{(r^\gamma+q^\gamma)^\frac{4}{\gamma}}-\frac{2m}{(r^\gamma+q^\gamma
)^\frac{3}{\gamma}}\bigg]\end{equation} where $P(r)$ is variable
pressure of central AdS space. It is easy to check that the
horizon equation $f(r_+)=0$ reads
\begin{equation}
\label{m11}1-\frac{2mr_+^2}{\left(r_+^\gamma+q^\gamma \right)
^{3/\gamma}}+\frac{q^2r_+^2}{\left(r_+^\gamma+q^\gamma \right)
^{4/\gamma}}=0
\end{equation} in which $r_+$ is exterior horizon radius and it can be shown that asymptotically
reaches to the following solutions
\begin{equation}r_+>>|q|,~~~~(r_+)_{1,2}=m\pm\sqrt{m^2-q^2},\end{equation}
and
\begin{equation}r_+<<|q|,~~~(r_+)_{1,2}=|q|\sqrt{\frac{q}{2m-q}}.\end{equation} The horizon equation (\ref{m11}) can be rewritten as the
enthalpy equation of the ABG black
 hole such that
\begin{equation}\label{m}
m=H=U+PV.
\end{equation}
Here we call $m=H$ to be enthalpy and $V$ and $U$ are
thermodynamic volume and internal energy respectively  with
following forms.
\begin{equation}\label{V}V(r_+)=\frac{4\pi}{3}(r_+^\gamma+q^\gamma)^\frac{3}{\gamma},~~~U(r_+)=\frac{q^2}{2(r_+^\gamma+q^\gamma)^\frac{1}{\gamma}}
\end{equation}
By looking at the equation (\ref{V}) one can infer that for this
black hole the thermodynamic volume has different form with
respect to its geometrical volume while for some of black hole
they have same forms. In fact thermodynamic volume of a black hole
system is conjugate quantity for the pressure in the black hole
equation of state and it is determined by the first law of the
black hole thermodynamics $TdS=dU+PdV.$ To study thermodynamics of
the ABG black hole we need to calculate its Hawking temperature
which is given versus the surface gravity on the exterior horizon
$r_+$. By substituting (\ref{m}) and (\ref{V}) into the metric
potential (\ref{f1}) and by calculating the surface gravity
$f^{\prime}(r_+)$ we obtain the Hawking temperature for the ABG
black hole as follows.
\begin{equation} \label{T}
T(r_+)=vP+F(v)
\end{equation}
where $F(v)$ is defined by specific volume $v$ for which we
defined
\begin{equation}\label{v}v=\frac{r_+}{3}\bigg(\frac{r_+^\gamma-2q^\gamma}{r_+^\gamma+q^\gamma}\bigg),~~
~F(v)=-\frac{q^2r_{+}^{1+\gamma}}{8\pi(r_+^\gamma+q^\gamma)^{1+\frac{4}{\gamma}}}.\end{equation}
In the next section we  investigate possibility of thermodynamic
phase transitions.
\section{Thermodynamic phase transitions}
The equations (\ref{T}) with (\ref{v}) are parametric forms for
the ABG black hole equation of state.
 To obtain a single closed form we must separate scales of the ABG black hole to small and large black holes which by regarding positivity
 condition on the specific volume the equation (\ref{v}) gives us
 \begin{equation}\label{41}\bigg|\frac{r_+}{q}\bigg|\leq 2^\frac{1}{\gamma}\to v\approx-\frac{r_+}{3},~~~
 T\approx Pv-\frac{1}{8\pi q}\frac{\big(\frac{-3v}{q}\big)^\gamma}{\big[
 1+\big(\frac{-3v}{q}\big)^\gamma\big]^{1+\frac{4}{\gamma}}}\end{equation} for small scale ABG black holes and
 \begin{equation}\label{42}\bigg|\frac{r_+}{q}\bigg|> 2^\frac{1}{\gamma}\to v\approx\frac{r_+}{3},~~~T\approx
 Pv-\frac{1}{8\pi q}\frac{\big(\frac{3v}{q}\big)^\gamma}{\big[
 1+\big(\frac{3v}{q}\big)^\gamma\big]^{1+\frac{4}{\gamma}}}\end{equation} for large scale ABG black holes.
For simplicity of investigation of the phase transition we set
ansatz $q=-3$ and $q=3$ for the equations (\ref{41}) and
(\ref{42}) respectively such that
 \begin{equation}\label{43}T\approx Pv+\frac{1}{24\pi}\frac{v^\gamma}{(
 1+v^\gamma)^{1+\frac{4}{\gamma}}},~~~0\leq v\leq2^\frac{1}{\gamma}\end{equation}
 for small ABG black holes and
 \begin{equation}\label{44}T\approx Pv-\frac{1}{24\pi}\frac{v^\gamma}{(
 1+v^\gamma)^{1+\frac{4}{\gamma}}},~~~v>2^\frac{1}{\gamma}\end{equation}
 for large ABG black holes respectively.
Now that, we are in position to find the critical thermodynamic
variables of the system through solving below equations.
\begin{equation}\label{crit}
\frac{\partial T}{\partial v}\Big|_{P}=0, ~~~ \frac{\partial^2
T}{\partial v^2}\Big|_{P}=0
\end{equation}
which by substituting (\ref{43}) and (\ref{44}) we obtain
parametric forms of the critical points $(v_c,P_c,T_c)$ in phase
space such that
\begin{equation}\label{46}(v_c^\pm)_{large}=
(v_c^{\pm})_{small}=\bigg[\frac{\gamma^2+13\gamma-4\pm\sqrt{\gamma^4+26\gamma^3+81\gamma^2-24\gamma+16}}{40}\bigg]^\frac{1}{\gamma},
\end{equation}
\begin{equation}\label{47}-(P_c^\pm)_{large}=(P_c^{\pm})_{small}=\frac{(v_c^{\pm})^{\gamma-1}[4(v_c^\pm)^\gamma-\gamma]}{24\pi[1+(v_c^{\pm})^\gamma]^{2+
\frac{4}{\gamma}}},\end{equation}\begin{equation}\label{48}-(T_c^\pm)_{large}=(T_c^{\pm})_{small}=\frac{(v_c^{\pm})^{\gamma-1}[5(v_c^\pm)^\gamma+1
-\gamma]}{24\pi[1+(v_c^{\pm})^\gamma]^{2+\frac{4}{\gamma}}}\end{equation}
and
\begin{equation}\label{sc3}P_{\gamma}=P(2^\frac{1}{\gamma}),~~~T_{\gamma}=T(2^\frac{1}{\gamma}).\end{equation}
We collected some numeric values for the critical pressures and
critical temperatures for different values of the $\gamma$
parameter in the table 2. At the first step, we should determine
numeric values for $\gamma$ parameter for which the phase
transitions may possible happens for small and large ABG black
holes. To do so we look diagrams of the critical specific volumes
(\ref{46}) given in the figure 1 for which we see a local maximum
point just for diagram of $v_c^+$ but not for $v_c^-.$ Furthermore
we can obtain
\begin{equation}\lim_{\gamma\to\pm\infty}v_{c}^\pm(\gamma)=1,~~~v_c^{\pm}(0)\to0.\end{equation}
 To study phase transition of this ABG black hole we must be select a suitable numeric value for $\gamma$ parameter given in the
 figure 1-a. As a suitable sample we select numeric value for $\gamma$ by solving the equation
 $\frac{dv_c^+}{d\gamma}=0$ for large ABG black hole such that $\gamma=4.858260939.$
By substituting $\gamma=4.858260939$ one can obtain
$v_c^+=1.323138639$ for specific volume  of largest ABG black hole
and $v_c^-=0.7457595729$ for specific volume of smaller one ABG
black hole. We say  these are large and small ABG black holes
because for this particular value of $\gamma$ the corresponding
limit specific volume reads $2^\frac{1}{\gamma}=1.153353666$ for
which the inequality condition $v_c^-<2^\frac{1}{\gamma}$ and
$v_c^+>2^{\frac{1}{\gamma}}$ is established. We collect all
components of the critical points given by (\ref{46}), (\ref{47})
and (\ref{48}) for this particular $\gamma$ value in the table 1
and corresponding critical pressures and critical temperatures are
collected in the table 2. Also we use (\ref{43}) and (\ref{44}) to
plot the P-V diagrams at constant temperatures in the figures 1-b
and 1-c for small (Large) ABG black hole with specific volume
$v^-_c(v_c^+)$ which are given in the table 1.  By looking at
these diagrams one can infer that 1-b shows a Van der Waals phase
transition at positive temperatures higher than the critical
(negative) temperature $T>T_c^+$ while it is in unstable state
thermally at below of the critical temperature $T\leq T_c^+$ and
may reaches from AdS $(P>0)$ to dS $(P<0)$. However by looking at
the figure 1-c we understand that the large ABG black hole is
always in the AdS phase $(P>0)$ but it participates in the small
to large black hole phase transition (a Van der Waals type) at
temperatures below the critical one $T\leq T_c^-.$ Also the
diagram shows that by raising the specific volume $v$ the black
hole passes from a local maximum pressure and then can be
participates in the Hawking-Page phase transition by making up as
a gas form with infinite specific volume.  Numeric values of the
critical points for choices
$\gamma=\{1,0.9,0.8,0.7,0.6,0.4,0.3,0.2,0\}$ are not obtained real
finite numbers. Mathematica software generates as imaginary
complex numbers or undetermined and so they are not shown in the
tables 1 and 2. Also we checked numerical solutions of the
critical points for negative $\gamma$ and obtained some complex
imaginary numbers again for $-22<\gamma<0$ but there are obtained
some real valued critical specific volume for $\gamma<-23$ for
which $1>2^\frac{1} {\gamma}>v_c^->v_c^+>0$ and so we do not
consider them again in this paper because they describe small ABG
black holes which we considered via positive values for $\gamma$
parameter. By substituting the internal energy and the
thermodynamic volume given by the equations (\ref{V}) and the
Hawking  temperature (\ref{T}) into the first law of the black
hole thermodynamics $TdS=dU+PdV,$ we obtain integral equation for
the Bekenstein entropy of this kind of the ABG black hole such
that
\begin{equation}S=12\pi q^2\int^{v_c^\pm} J(v)dv,\end{equation} in
which we defined
 \begin{equation}J(v)=\frac{v^{\gamma-1}(1+v^\gamma)^{\frac{3}{\gamma}}[8\pi Pq^2(1+v^\gamma)^\frac{4}{\gamma}-1]}{8\pi Pq^2(v^\gamma
 -2)(1+v^\gamma)^\frac{4}{\gamma}-3v^{1+\gamma}}.\end{equation} The above integral equation has not an analytic
  closed form solution regretfully  and so to use it in drawing the Gibbs free energy  $G=m-TS$ we   should use
  numerical calculations (diagrams for  critical $J_c^{\pm}$ are plotted in figure 1-b ). But we can use variations of the entropy $J(v)$ to obtain a closed form for heat capacity at
  constant pressure. This is done by according to ordinary thermodynamics systems where the heat capacity at constant
 pressure is defined by $C_P=T\big(\frac{\partial S}{\partial T}\big)_P.$ Thus we substitute the above entropy equation and temperatures
 (\ref{43}) and (\ref{44}) and the charge value $q=\pm3$
 to obtain
 \begin{equation}C_P(v)
 =\frac{108\pi v^\gamma(1+v^\gamma)^{1+\frac{3}{\gamma}}[24\pi
P(1+v^\gamma)^{1+\frac{4}{\gamma}}+\epsilon v^{\gamma-1}][72\pi
P(1+v^\gamma)^\frac{4}{\gamma}-1]}{[24\pi
P(1+v^\gamma)^{2+\frac{4}{\gamma}}+\epsilon
v^{\gamma-1}(\gamma-4v^\gamma)][72\pi P(v^\gamma
 -2)(1+v^\gamma)^\frac{4}{\gamma}-3v^{1+\gamma}]}\end{equation} in which
$\epsilon=+1(-1)$ corresponds to small (large) black holes.
Diagram of the above heat capacity is plotted vs $v$ for small and
large ABG black holes in the figure 2-d for critical point given
in the table 1 for $\gamma=4.86.$
\section{Conclusion}
By extracting a formal variable cosmological parameter from the
generalized nonsingular ABG black hole where its central region
behaves same as dS/AdS space we can obtain a suitable equation of
state in which the assumed variable cosmological parameter plays
same as the pressure thermodynamic coordinate. Positivity
condition of the specific thermodynamic volume of the black hole
system restrict us to separate negative (positive) sign for the
magnetic charge to produce some acceptable equation of states
which are corresponded to some small (large) scale generalized ABG
black hole. By plotting isothermal P-V curves at constant
temperature and also by plotting the heat capacity at constant
pressure for both of small and large scale ABG black hole we
obtained situations for these diagrams where the black hole can be
participate in one of two possible phase transitions called as (a)
small to large phase transition or (b) Hawking-Page transition. In
the former case an unstable quantum black hole changes its scale
after evaporation but in the later one a black hole evaporates
full and reaches to a gas with infinite volume. It is same as Van
der Waals gas/fliud phase transition in the ordinary thermodynamic
system. As extension of this work we like to study other
thermodynamic behavior of the modified ABG black hole such as heat
engine and Joule-Thomson expansion and etc.
\section{  Acknowledgment }  A preprint of this work has
previously been published.\cite{0}.\\
The authors are grateful to the editor and anonymous referees for
their valuable comments and suggestions to improve the paper.

\newpage
\begin{figure}[ht] \centering  \subfigure[{}]{\label{10001}
\includegraphics[width=0.4\textwidth]{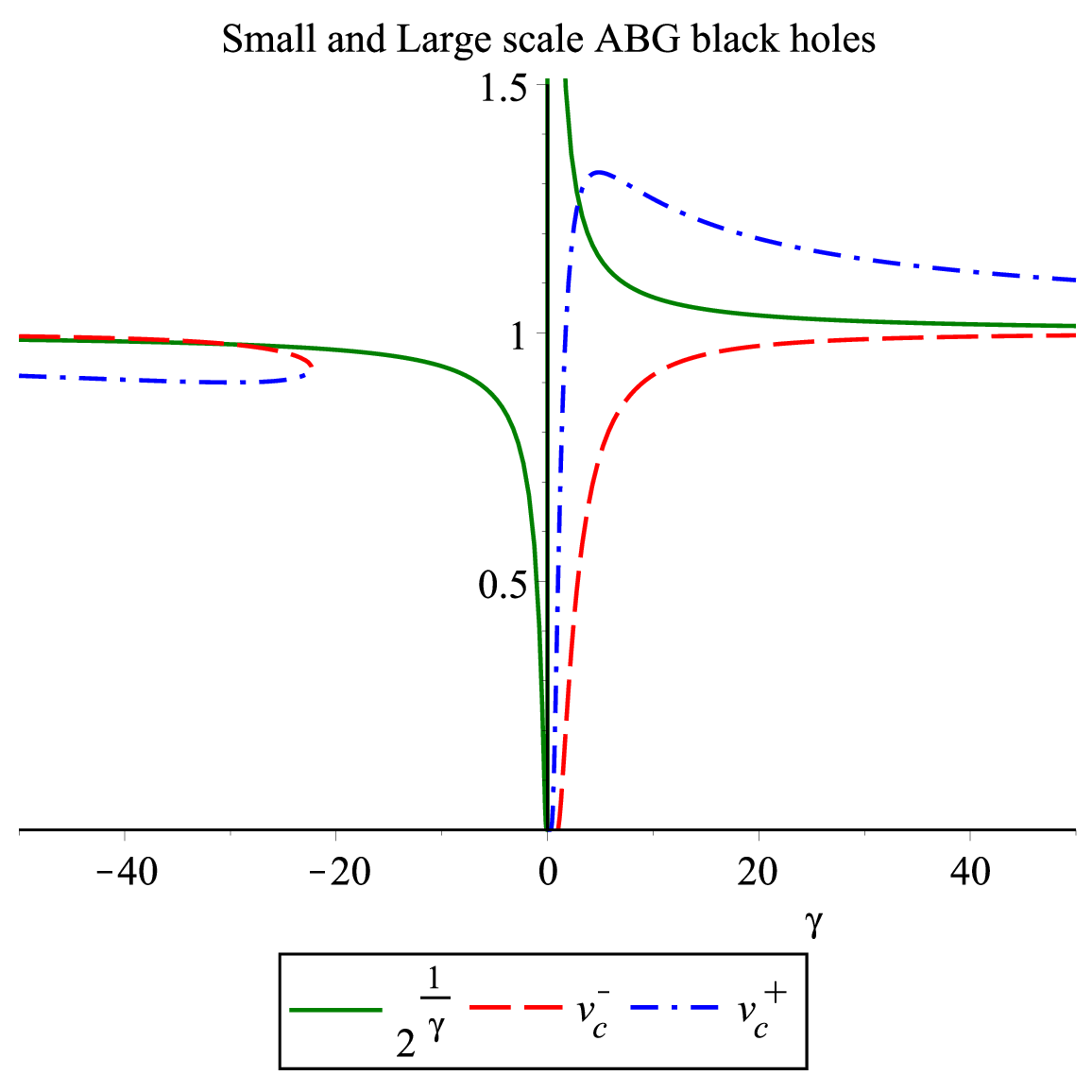}}
\hspace{3mm}\subfigure[{}]{\label{23401}
\includegraphics[width=0.4\textwidth]{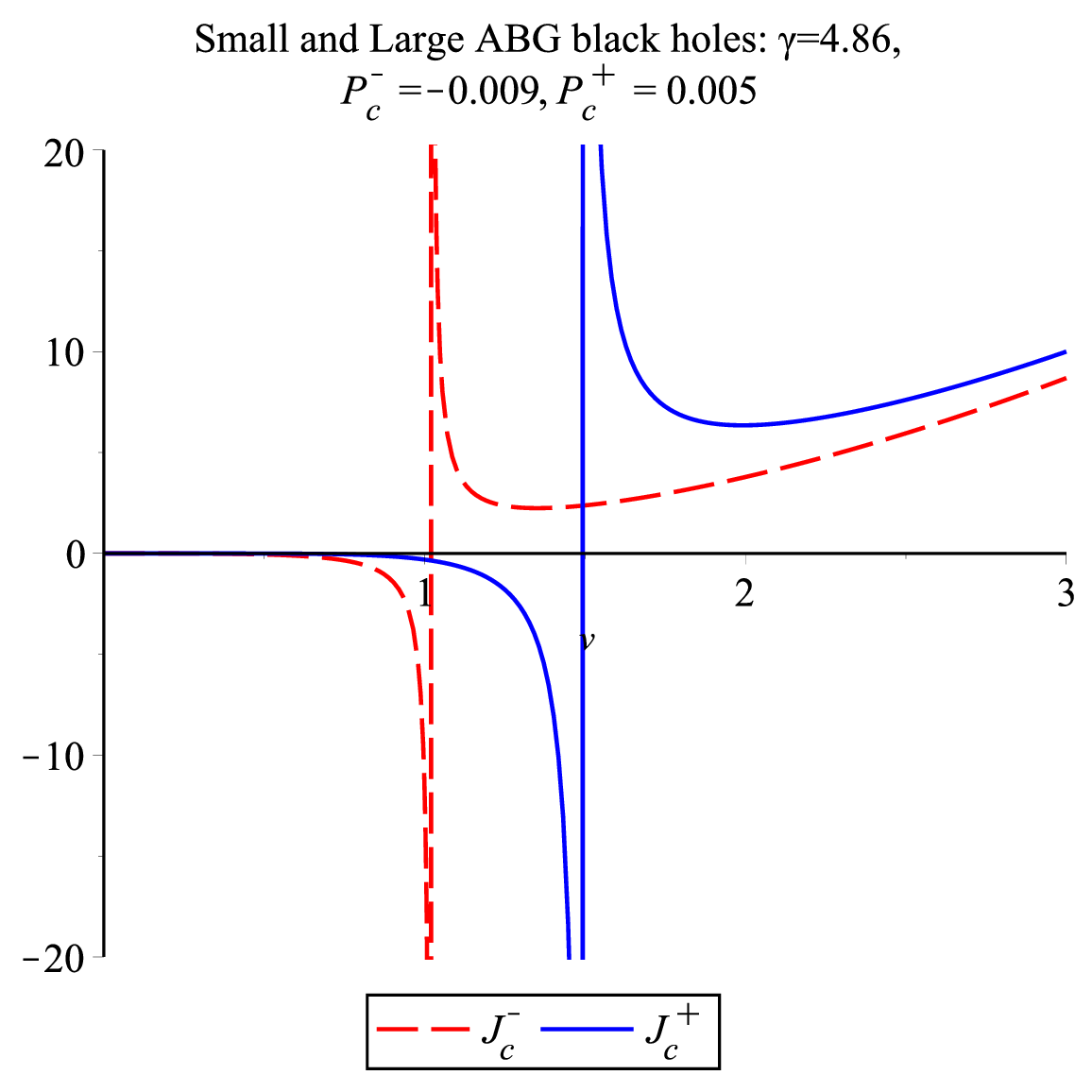}}
\hspace{3mm}  \caption{(a) Diagrams of the critical specific
volume of small and large generalized ABG black holes vs $\gamma$
and (b) Variation of entropy per specific volume of small
$J^-=\frac{\Delta S}{\Delta v^-}$ and large $J^+=\frac{\Delta
S}{\Delta v^+}$ generalized ABG black holes }
\end{figure}
\begin{figure}[ht]
\centering  \subfigure[{}]{\label{10001}
\includegraphics[width=0.4\textwidth]{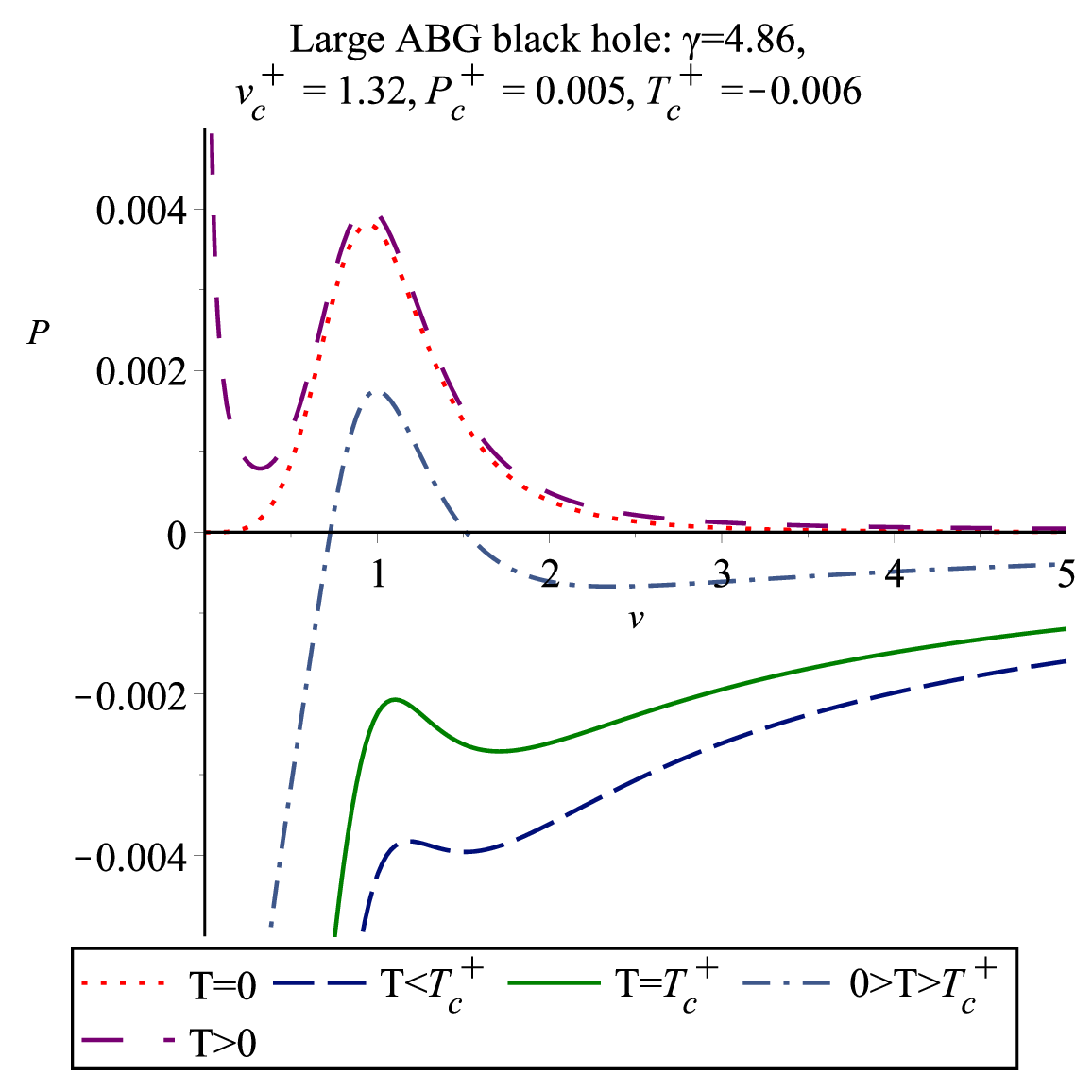}}
\hspace{3mm}\subfigure[{}]{\label{23401}
\includegraphics[width=0.33\textwidth]{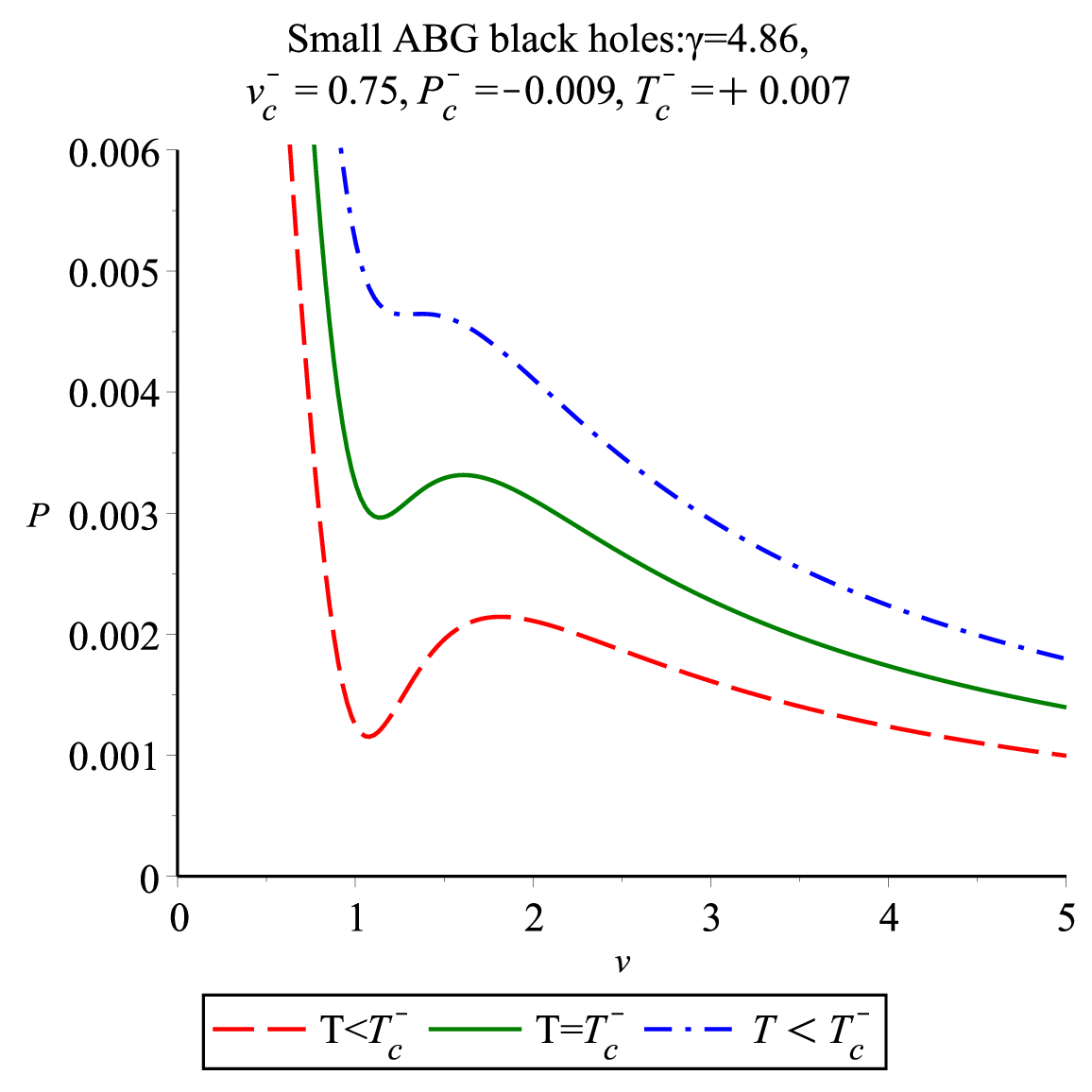}}
\hspace{3mm} \subfigure[{}]{\label{s01}
\includegraphics[width=0.33\textwidth]{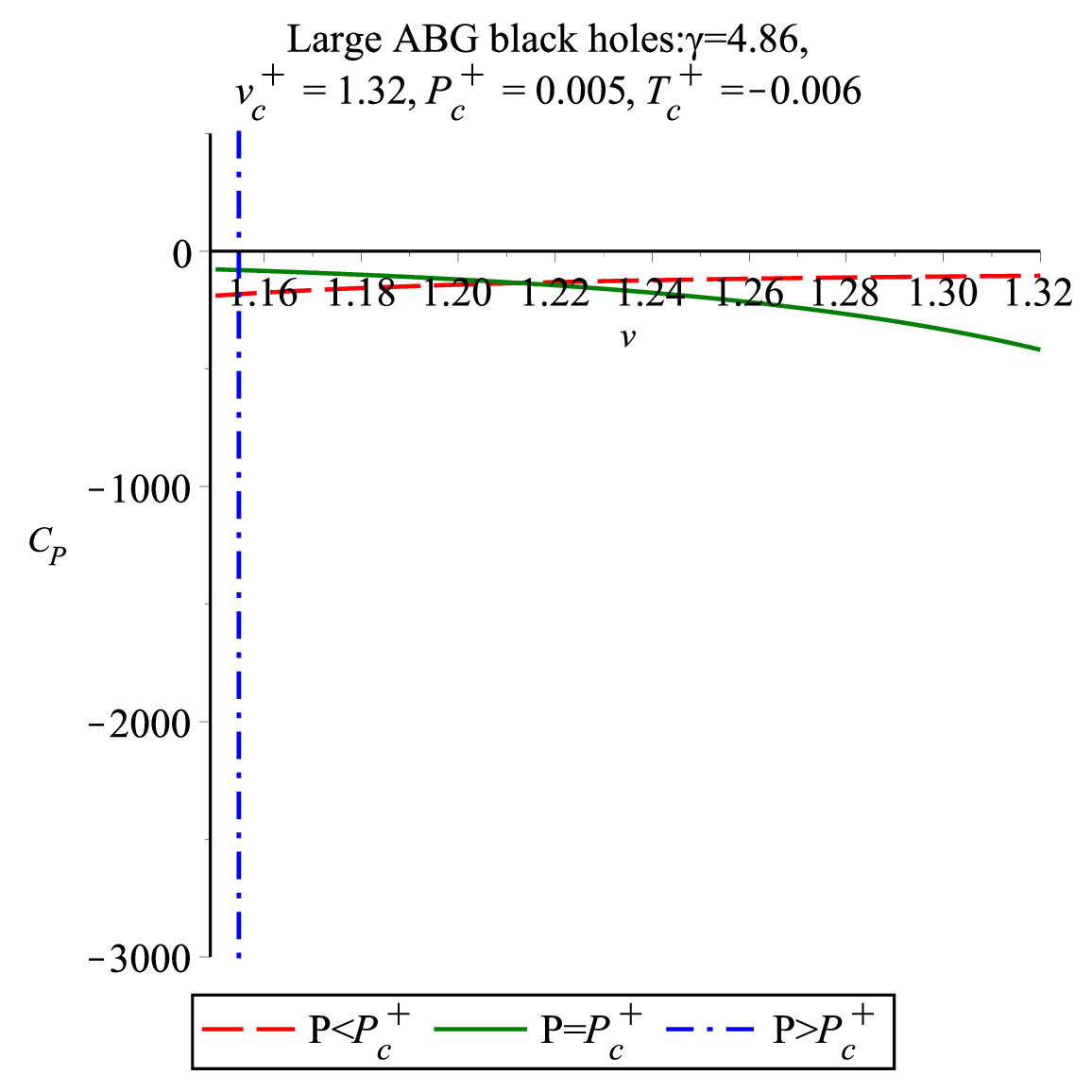}}
\hspace{3mm} \subfigure[{}]{\label{a91}
\includegraphics[width=0.33\textwidth]{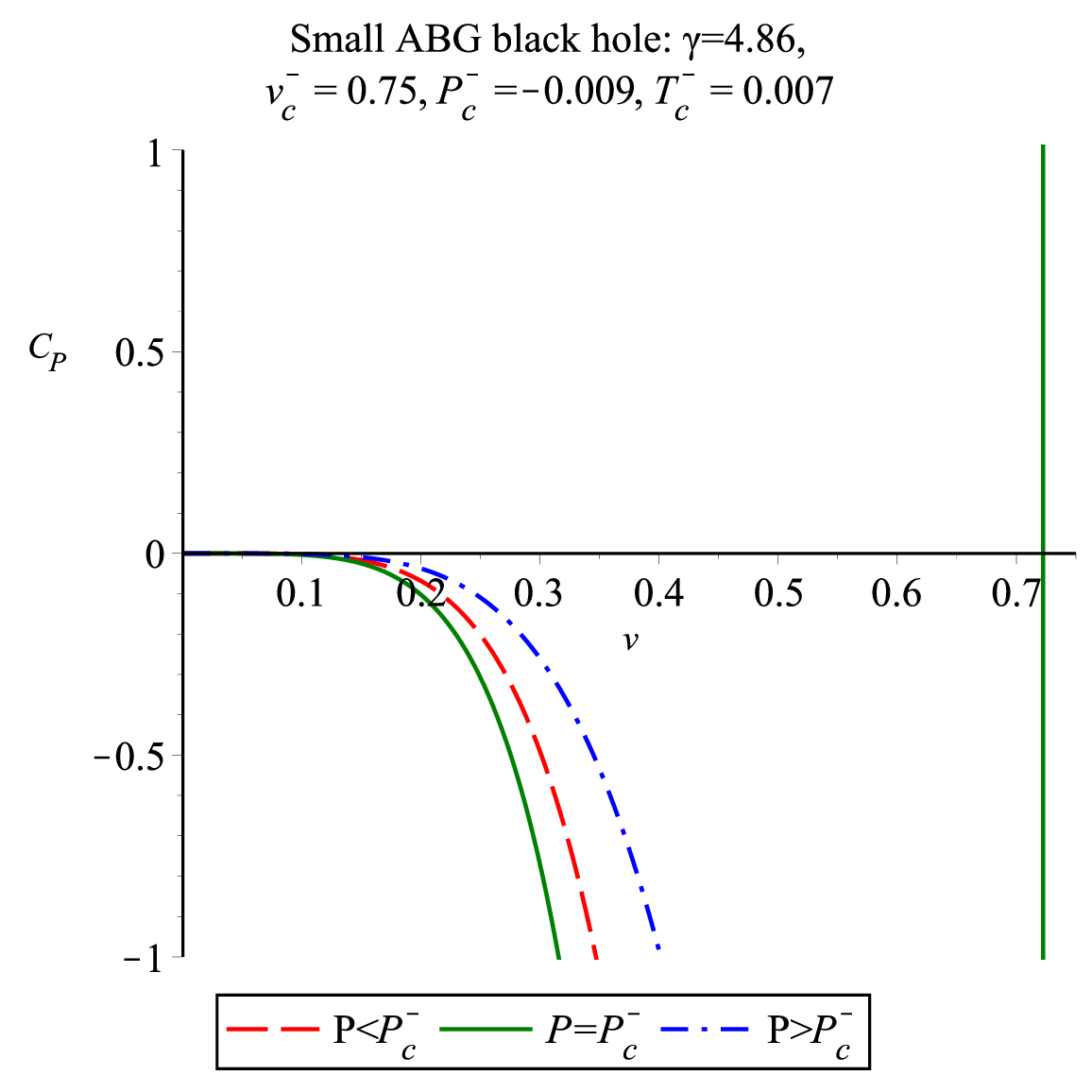}}
\hspace{3mm}  \caption{ (a) and (b): P-V diagrams at constant
temperatures for large and small generalized ABG black hole
respectively, (c) and (d): Heat capacity diagrams at constant
pressure for large and small generalized ABG black holes
respectively}
\end{figure}
\newpage
\begin{center}Table 1: Critical
specific volume for large $v_c^+$ and small $v_c^-$ generalized
ABG black holes
\end{center}
\begin{center}
\begin{tabular}{|c|c|c|c|}
\hline
$\gamma$ &$2^\frac{1}{\gamma}$ &$v_c^-$ & $v_c^+$  \\
\hline
$\pm\infty$ & 1 & 1 & 1 \\
4.86 &1.15 &0.75 & 1.32  \\
4.8& 1.15535& 0.741427& 1.32311\\
4.7& 1.15891& 0.733726&  1.32295\\
4.6& 1.16263& 0.725675& 1.32263\\
4.5& 1.16653& 0.717254&  1.32212\\
4.4& 1.17062& 0.708439& 1.32141\\
4.3& 1.17492& 0.699207&  1.32048\\
4.2& 1.17943& 0.689532& 1.3193\\
4.1& 1.18419& 0.679386&  1.31784\\
4.0& 1.18921& 0.66874& 1.31607\\
3.9& 1.1945& 0.657562&  1.31397\\
3.8& 1.2001& 0.645819& 1.31148\\
3.7& 1.20603& 0.633474&  1.30856\\
3.6& 1.21233& 0.620488& 1.30517\\
3.5& 1.21901& 0.60682&  1.30125\\
3.4& 1.22613& 0.592427& 1.29674\\
3.2& 1.24186& 0.561278&  1.28563\\
3.1& 1.25057& 0.54442& 1.27886\\
3.0& 1.25992& 0.526637&  1.27115\\
2.9&   1.27& 0.507872& 1.26236\\
2.8& 1.28089& 0.48807& 1.25237\\
2.7& 1.29268& 0.467174&  1.24101\\
2.6& 1.30551& 0.445129& 1.22809\\
2.5& 1.31951& 0.421882&  1.21341\\
2.4& 1.33484& 0.39739& 1.19673\\
2.3& 1.35171& 0.371616&  1.17775\\
2.2& 1.37035& 0.344542& 1.15616\\
2.1& 1.39107& 0.316172& 1.13158\\
2.0& 1.41421& 0.286547&  1.10358\\
1.9&   1.44025& 0.255752& 1.07166\\
1.8& 1.46973& 0.223943& 1.03526\\
1.7& 1.50341& 0.191364&   0.993727\\
1.6& 1.54221& 0.158386&  0.946364\\
1.5& 1.5874& 0.125544&  0.892396\\
1.4& 1.64067& 0.0935867&  0.831023\\
1.3& 1.70436& 0.0635304&  0.761483\\
1.2& 1.7818& 0.0367104&  0.683169\\
1.1& 1.87786& 0.0148133&  0.595845\\
0.5& 4.0& 0.00390625&  0.04\\
0.1&   1024.& $1.26329\times10^{-8}$& $2.69539\times10^{-16}$\\
\hline
\end{tabular}
\end{center}
\begin{center}Table2: Critical pressure and temperature for large $(P_c^+,T_c^+)$ and
 small $(P_c^-,T_c^-)$  generalized ABG black holes \end{center}
\begin{center}
\begin{tabular}{|c|c|c|c|c|c|c|c|}
\hline
$\gamma$  & $P_c^-$ & $P_c^+$ & $T_c^-$ & $T_c^+$ \\
\hline
$\pm\infty$  & 0.053 & 0.053 & 0.066 &0.066\\
4.86  & -0.009 & 0.005 & 0.007 & -0.006 \\
4.8& -0.00894586& 0.00466126&  $5.21225\times10^{-6}$& $2.40448\times10^{-6}$\\
4.7&  -0.00873715&0.00455079&$4.74355\times10^{-6}$&  $2.1582\times10^{-6}$\\
4.6&  -0.00853071&  0.0044403& $4.29998\times10^{-6}$& $1.92837\times10^{-6}$\\
4.5&  -0.00832665& 0.00432984&
$3.88149\times10^{-6}$&  $1.71469\times10^{-6}$\\
4.4&  -0.00812505&  0.00421943& $3.48802\times10^{-6}$& $1.5168\times10^{-6}$\\
4.3&  -0.00792601& 0.00410912&
$3.11942\times10^{-6}$&  $1.33433\times10^{-6}$\\
4.2&  -0.00772965&  0.00399893& $2.77545\times10^{-6}$& $1.16684\times10^{-6}$\\
4.1&  -0.00753608& 0.00388892&
$2.45584\times10^{-6}$&  $1.01386\times10^{-6}$\\
4.0&  -0.00734544&  0.00377911& $2.16019\times10^{-6}$& $8.74879\times10^{-7}$\\
3.9&  -0.00715786& 0.00366957&
$1.88807\times10^{-6}$&  $7.49348\times10^{-7}$\\
3.8&  -0.0069735&  0.00356033& $1.63892\times10^{-6}$& $6.36675\times10^{-7}$\\
3.7&  -0.00679254& 0.00345145&
$1.41213\times10^{-6}$&  $5.36231\times10^{-7}$\\
3.6&  -0.00661518&  0.00334299& $1.20698\times10^{-6}$& $4.47353\times10^{-7}$\\
3.5&  -0.00644162& 0.00323501&
$1.02268\times10^{-6}$&  $3.69348\times10^{-7}$\\
3.4&  -0.00627213&  0.00312758& $8.58331\times10^{-7}$& $3.01494\times10^{-7}$ \\
3.2& -0.00594647& 0.00291466&
$5.85581\times10^{-7}$&  $1.93243\times10^{-7}$\\
3.1&  -0.005791&  0.00280933& $4.75014\times10^{-7}$& $1.51311\times10^{-7}$\\
3.0&  -0.00564099& 0.00270487& $3.801\times10^{-7}$& $1.16475\times10^{-7}$\\
2.9&    -0.00549696& 0.00260139&
$2.99601\times10^{-7}$&  $8.79622\times10^{-8}$\\
2.8&  -0.0053595&  0.002499& $2.32241\times10^{-7}$& $6.50117\times10^{-8}$\\
2.7&  -0.00522933&
0.00239781&$1.76714\times10^{-7}$&  $4.68846\times10^{-8}$\\
2.6&  -0.00510734&  0.00229797& $1.31704\times10^{-7}$& $3.28717\times10^{-8}$\\
2.5&  -0.00499461&
0.00219962&$9.59015\times10^{-8}$&  $2.23033\times10^{-8}$\\
2.4&  -0.0048925&  0.00210292& $6.80237\times10^{-8}$& $1.4557\times10^{-8}$\\
2.3&  -0.0048027& 0.00200808& $4.6836\times10^{-8}$& $9.06631\times10^{-9}$\\
2.2&   -0.00472744&
0.00191531&$3.11721\times10^{-8}$&   $5.32651\times10^{-9}$\\
2.1&  -0.00466961&  0.00182485& $1.99541\times10^{-8}$& $2.89986\times10^{-9}$\\
2.0&  -0.00463312& 0.001737& $1.22099\times10^{-8}$& $1.41809\times10^{-9}$\\
1.9&    -0.00462331&
0.00165211&$7.08794\times10^{-9}$&   $5.8254\times10^{-10}$\\
1.8&  -0.00464778&  0.00157061&
$3.86669\times10^{-9}$& $1.61493\times10^{-10}$\\
1.7&  -0.00471762& 0.001493&  $1.95847\times10^{-9}$& $-1.52291\times10^{-11}$\\
1.6&  -0.00484969& 0.00141997&  $9.06514\times10^{-10}$& $-6.42181\times10^{-11}$\\
1.5&  -0.00507089& 0.00135239&  $3.75326\times10^{-10}$& $-5.82997\times10^{-11}$\\
1.4&  -0.00542679& 0.00129145&  $1.34828\times10^{-10}$& $-3.73485\times10^{-11}$\\
1.3&  -0.00600098& 0.00123884&  $4.00938\times10^{-11}$& $-1.8947\times10^{-11}$\\
1.2&  -0.00696523& 0.00119706&  $9.07189\times10^{-12}$& $-7.7284\times10^{-12}$\\
1.1&  -0.00874552& 0.00117005&  $1.26329\times10^{-12}$& $-2.48105\times10^{-12}$\\
0.5&  -0.0289341&  0.00321305& $-6.99671\times10^{-20}$& $-3.26406\times10^{-20}$\\
0.1&   169.214&  $4.73581\times10^9$& $-3.67174\times10^{-82}$& $-4.52954\times10^{-80}$\\
\hline
\end{tabular}
\end{center}
\end{document}